\documentclass[aps,prl,twocolumn,superscriptaddress]{revtex4}
\usepackage{graphicx}
\usepackage{amssymb}
\usepackage{amsmath}
\usepackage{graphicx}
\usepackage{epsfig}
\usepackage{array}
\usepackage{color}
\usepackage{multirow}

\begin{document}

\title{Universal behavior in the Nernst effect of heavy fermion materials}

\author{Yi-feng Yang}
\email[]{yifeng@iphy.ac.cn}
\affiliation{Beijing National Laboratory for Condensed Matter Physics and Institute of Physics, Chinese Academy of Sciences, Beijing 100190, China}
\affiliation{University of Chinese Academy of Sciences, Beijing 100049, China}
\affiliation{Songshan Lake Materials Laboratory, Dongguan, Guangdong 523808, China}

\date{\today}

\begin{abstract}
We report the observation of a universal scaling of the Nernst coefficient over a wide intermediate temperature range in heavy fermion materials, including the superconductors CeCu$_2$Si$_2$, CeCoIn$_5$, and Ce$_2$PdIn$_8$, the ferromagnetic Kondo lattice Ce$_3$RhSi$_3$, the nonmagnetic CeRu$_2$Si$_2$, the intermediate valent YbAl$_3$, and the hidden order compound URu$_2$Si$_2$, that cover a broad spectrum of heavy fermion materials with different crystal, valence and ground state properties. The scaling formula follows exactly the magnetic susceptibility of emergent heavy quasiparticles as predicted in the two-fluid model. We give a tentative explanation of the scaling based on the skew scattering mechanism and the Boltzmann picture and argue that the Nernst effect is produced by the asymmetry of the quasiparticle density of states rather than that of the scattering rate. In URu$_2$Si$_2$, the giant Nernst signal in the hidden order phase is also found to follow the predicted scaling, indicating the potential involvement of hybridization physics. Our work suggests a candidate unified scenario for the Nernst effect in heavy fermion materials.
\end{abstract}

\maketitle

\section{I. Introduction}

The Nernst effect, a thermoelectric analog of the Hall effect, measures the transverse electric field induced by a longitudinal thermal gradient under a perpendicular magnetic field. It was first observed in elemental bismuth \cite{Nernst1886,Behnia2007} but is often extremely small (nV/KT)  in simple metals due to the Sondheimer cancellation \cite{Sondheimer1948}. A large Nernst signal of the order of $\mu$V/KT may be realized in anisotropic metals with nonspherical Fermi surfaces \cite{Clayhold1996} or multiband systems where the presence of both electron and hole carriers can give rise to a large ambipolar contribution \cite{Bel2003}. But unlike those in semiconductors and semimetals \cite{Delves1965}, the Nernst effect in correlated systems has not been paid much attention until an anomalously enhanced signal was discovered and attributed to the movement of vortices or vortexlike excitations in the pseudogap phase of the underdoped La$_{2-x}$Sr$_x$CuO$_4$ \cite{Xu2000}. This resembles that of conventional or high-$T_c$ superconductors \cite{Huebener1969,Hagen1990,Ri1994} and immediately stimulated intensive interest in the community of correlated electrons \cite{Wang2001,Capan2002,Wang2002,Ussishkin2002,Wang2003,Capan2003,Wang2006,Zhu2008,Bel2004}.

In heavy fermion materials, large Nernst signals have since been reported and explained by various scenarios besides vortex motion \cite{Behnia2009}. Magnetic fluctuations, the proximity of a quantum critical point, or some collective modes have all been considered as possible origins, but a generic understanding has not been achieved, thus preventing the development of a unified theory that could potentially cover the rich variety of experimental observations. In this work, we report an unexpected discovery of universality in the Nernst effect via a systematic examination of existing data in a number of prototypical heavy fermion compounds. Independent of material details, we find that the Nernst coefficients in all  compounds exhibit a universal scaling over a wide intermediate temperature range, which follows exactly the predicted magnetic susceptibility of emergent heavy quasiparticles in the two-fluid model \cite{Yang2016}. This motivates us to develop a tentative explanation based on the skew scattering mechanism and relate the Nernst coefficient with the asymmetry of the quasiparticle density of states at the Fermi energy. Interestingly, the same scaling is also observed in the hidden order phase of URu$_2$Si$_2$, indicating potential involvement of hybridization physics in this mysterious state. Our observation provides an important clue for developing a microscopic theory of the Nernst effect in heavy fermion materials.

\section{II. Analyses of experimental data}

\begin{figure}[t]
\begin{center}
\includegraphics[width=0.5\textwidth]{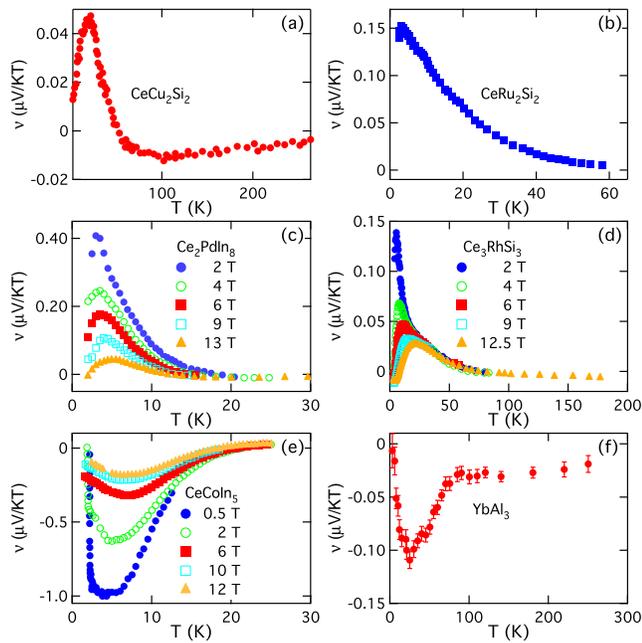}
\caption{Collection and comparison of the Nernst coefficients ($\nu$) in a number of prototypical heavy fermion compounds including (a) CeCu$_2$Si$_2$ \cite{Sun2013}, (b) CeRu$_2$Si$_2$ \cite{Behnia2009}, (c) Ce$_2$PdIn$_8$ \cite{Matusiak2011}, (d) Ce$_3$RhSi$_3$ \cite{Matusiak2013}, (e) CeCoIn$_5$ \cite{Bel2004}, and (f) YbAl$_3$ \cite{Wei2015}. The data were reproduced from the literature, possibly with different sign conventions.}
\label{fig1}
\end{center}
\end{figure}

Figure \ref{fig1} collects and compares the Nernst data reproduced from the literature for the heavy fermion superconductors CeCu$_2$Si$_2$ \cite{Sun2013}, CeCoIn$_5$ \cite{Bel2004,Sheikin2006}, and Ce$_2$PdIn$_8$ \cite{Matusiak2011}, the ferromagnetic Kondo lattice Ce$_3$RhSi$_3$ \cite{Matusiak2013}, the nonmagnetic CeRu$_2$Si$_2$ \cite{Behnia2009}, and the intermediate valence compound YbAl$_3$ \cite{Wei2015}. These cover a broad spectrum of heavy fermion systems with different crystal, valence and ground state properties. The magnitude of their Nernst coefficients also varies over two orders of magnitude and has been given different explanations. In CeCu$_2$Si$_2$, an enhanced Nernst coefficient was observed to correlate with the thermopower and ascribed to the asymmetric (or skew) Kondo scattering \cite{Sun2013}. In YbAl$_3$, the correlation is absent,  possibly due to its intermediate valence. A finite Nernst signal was observed at high temperatures and attributed to other scattering processes such as acoustic phonons \cite{Wei2015}. In the ferromagnetic Kondo lattice Ce$_3$RhSi$_3$ \cite{Matusiak2013}, the Nernst coefficient shows a similar temperature dependence as the Hall coefficient, most likely originating from the skew scattering \cite{Fert1987}, but the thermopower behaves very differently and cannot be explained by available theories. In CeCoIn$_5$, the maximal Nernst coefficient can reach 1 $\mu$V/kT at low field, much larger than the residual value in cuprates above $T_c$ \cite{Bel2004}. Various origins have been assigned such as antiferromagnetic fluctuations \cite{Izawa2007}, long-range phase coherence \cite{Onose2007}, or unconventional density wave \cite{Dora2005}. Moreover, $\nu/T$ was found to be dramatically enhanced at 6 T, most probably due to  a suppressed Fermi energy induced by the proximity of a field-indued quantum critical point \cite{Izawa2007}. In Ce$_2$PdIn$_8$, the Nernst coefficient diverges logarithmically below 7 K, implying the presence of an underlying quantum critical point \cite{Matusiak2011}. At intermediate temperatures, it exhibits strong field and temperature dependence, possibly associated with an anisotropic scattering time due to antiferromagnetic fluctuations. No ambipolar enhancement was detected at high temperatures despite of its multiband electronic structure.

\begin{figure}[t]
\begin{center}
\includegraphics[width=0.5\textwidth]{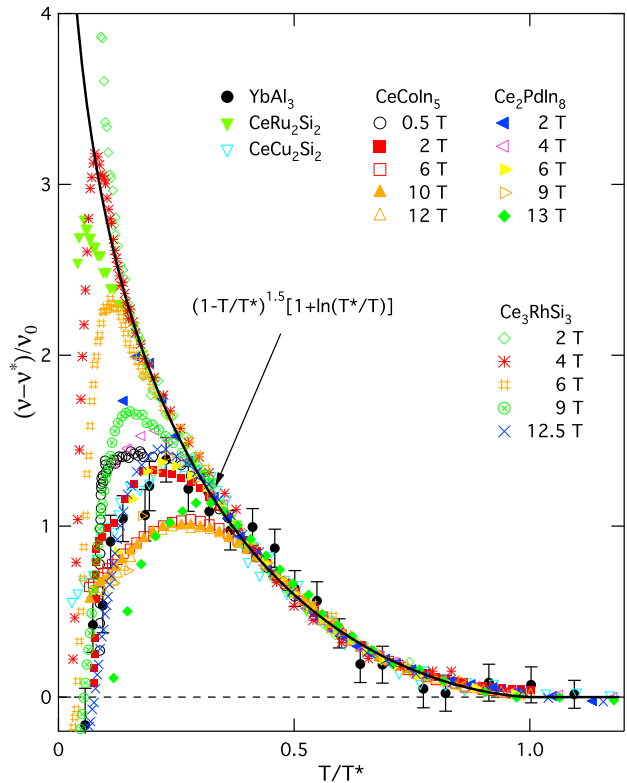}
\caption{Universal scaling of the normalized Nernst coefficients as a function of the dimensionless temperature for all data presented in Fig.~\ref{fig1}. The high temperature constants are subtracted. The solid line is the predicted scaling of emergent heavy quasiparticles in the two-fluid model \cite{Yang2008}.}
\label{fig2}
\end{center}
\end{figure}

Given such complexity, it seems unlikely to develop a unified picture. However, one may notice that the Nernst coefficients in all above compounds share certain kind of common features, namely a weak temperature dependency at high temperatures and a strong enhancement at intermediate temperatures, followed  by a rapid suppression as the temperature approaches zero. With increasing magnetic field, as shown for CeCoIn$_5$, Ce$_2$PdIn$_8$, and Ce$_3$RhSi$_3$, the Nernst coefficient is systematically suppressed and the peak shifts towards higher temperature, possibly following the increase of the Zeeman energy. This common trend suggests the possibility of a generic mechanism and motivates us to make a systematic comparison of existing experiments, which, quite unexpectedly, immediately reveals the presence of a true universality. As plotted in Fig.~\ref{fig2}, all data can be scaled to fall on a single curve over a wide intermediate temperature range. In most cases, the scaling works pretty well from $T^*$ down to the peak temperature. Moreover, it follows exactly the magnetic susceptibility (or density of states) of the emergent heavy quasiparticles predicted in the two-fluid model \cite{Yang2008},
\begin{equation}
\frac{\nu-\nu^*}{\nu_0}=\left(1-\frac{T}{T^*}\right)^{1.5}\left(1+\ln\frac{T^*}{T}\right),
\label{eq1}
\end{equation}
where $T^*$ is the onset temperature, $\nu^*$ is a temperature-independent constant from the high temperature background, and $\nu_0$ is a temperature-independent factor. The presence of these constants do not affect the overall temperature scaling. In heavy fermion materials, the high temperature localized $f$ electrons hybridize with conduction electrons and turn gradually itinerant with lowering temperature. The two-fluid model states that this delocalization process can be described phenomenologically by two coexisting fluids: an itinerant fluid of the emergent heavy quasiparticles and a spin liquid formed by the residual unhybridized moments \cite{Yang2016}. The model predicts that the itinerant fluid has a magnetic susceptibility of the above universal form, which has been confirmed in many experiments, in particular through the so-called Knight shift anomaly \cite{Nakatsuji2004,Curro2004,Yang2008}. At low temperatures, the scaling is typically interrupted by other orders or the Fermi liquid \cite{Yang2012,Shirer2012}.

\section{III. A phenomenological theory}

What is then the underlying mechanism? The above connection suggests a correlation between the Nernst signal and the magnetic susceptibility of the emergent heavy quasiparticles and motivates us to develop a tentative explanation based on the skew scattering mechanism. Quite generally, it has been known from the Boltzmann equation that the Nernst signal is connected with the Hall angle \cite{Behnia2009,Oganesyan2004},
\begin{equation}
\nu=-\frac{\pi^2k_B^2T}{3eH}\left.\frac{\partial\tan\Theta_H}{\partial\epsilon}\right|_{\epsilon_F},
\label{eq2}
\end{equation}
where $k_B$ is the Boltzmann constant, $e$ is the electron charge, $H$ is the magnetic field, and $\epsilon_F$ is the Fermi energy. The Hall angle is defined as $\tan\Theta_H=\rho_{xy}/\rho$, in which $\rho_{xy}$ and $\rho$ are the transverse and longitudinal resistivity, respectively. We have assumed energy dependency in all quantities. For the dominant skew scattering mechanism as confirmed in many heavy fermion materials \cite{Schoenes1988,Sakamoto2003,Chen2004,Sugawara2005,Paschen2005,Kohler2007}, the Hall coefficient is given by $R_H=R_0+r\rho\chi$, where $R_0$ is the normal contribution from background conduction electrons, $r$ is a constant prefactor, and $R_s=r\rho\chi$ is the skew scattering contribution in proportion to the magnetic susceptibility $\chi$ \cite{Fert1987}. Hence, the Nernst coefficient should also have two contributions: a background term, which is typically small as in normal metals, and a skew scattering term given by $\nu_s\propto T\partial \chi(\epsilon)/\partial\epsilon$ following the above Boltzmann picture. 

Now the two-fluid model states, $\chi=\chi_l+\chi_h$, where $\chi_h$ is the susceptibility of emergent heavy quasiparticles and $\chi_l$ is that of residual unhybridized moments \cite{Nakatsuji2004,Curro2004,Yang2008}. If the compound is in the Kondo limit, the local moments stay deep below the Fermi level so that $\chi_l$ must depend weakly on $\epsilon$. We have then $\partial \chi_l/\partial\epsilon|_{\epsilon_F} \approx 0$ and consequently a weak Nernst signal at high temperatures in the fully localized regime. This is indeed the case for all compounds in Fig.~\ref{fig1}. By contrast, the heavy electron spectral weight is always located near the Fermi energy and provides the major contribution, $\nu_s\propto \partial \chi_h(\epsilon)/\partial\epsilon |_{\epsilon_F}$. This is, unfortunately, still difficult to model in the lack of a full microscopic theory of the heavy fermion physics. To proceed, we consider the possibility of an $E/T$ (energy/temperature) scaling in the quasiparticle spectra, which has been observed in the scanning tunneling and angle-resolved photoemission spectroscopies of CeCoIn$_5$ \cite{Aynajian2012,Chen2017}. It is likely supported by the universal nature of the emergent heavy quasiparticles as predcited by the two-fluid model and examined in many experiments \cite{Yang2016}. One may then assume a generic form for the susceptibility (or density of states), $\chi_h(\epsilon)=\chi_h(T)g(x)$, where $\chi_h(T)$ gives the temperature dependence of the quasiparticle spectrum and $g(x)$ is a regular function of $x=(\epsilon-\epsilon_F)/T$ describing the shape of the spectrum. We have $\partial\chi_h(\epsilon)/\partial\epsilon|_{\epsilon_F}=-\chi_h(T)T^{-1}\partial g(x)/\partial x|_{x=0}$, which immediately yields the observed scaling $\nu_s \propto \chi_h(T)$. 

\section{IV. Discussion}

Two remarks are in order concerning this probably oversimplified derivation. First, the enhanced Nernst coefficient follows the emergence of itinerant heavy quasiparticles. Our fit yields $T^*\approx$ 110 K for YbAl$_3$, 59 K for CeRu$_2$Si$_2$, 78 K for CeCu$_2$Si$_2$, 98-107 K for Ce$_3$RhSi$_3$, 25 K for CeCoIn$_5$ and 16-18 K for Ce$_2$PdIn$_8$. These values agree well with those estimated from many other measurements for YbAl$_3$ ($120\pm10$ K), CeRu$_2$Si$_2$ ($60\pm10$ K), and CeCu$_2$Si$_2$ ($75\pm20$ K)  \cite{Yang2008b}. However, for CeCoIn$_5$, previous estimates give $T^*\approx40$ K along planar direction, which are twice larger than that from the Nernst fit. The same is also seen in Ce$_2$PdIn$_8$. This is very puzzling. It might be that the formula of $R_s$ fails below the coherence temperature. In fact, similar $\chi_h$ scaling has been observed in the Hall coefficient of CeCoIn$_5$ and Ce$_2$PdIn$_8$ \cite{Yang2013}. It is not clear if the discrepancy in these two compounds is accidental or demands an alternate explanation. At the moment, we cannot exclude other possibilities, but it seems hard to derive the scaling formula on a different basis. On the other hand, our Nernst data for CeCoIn$_5$ is only below 25 K. It will be interesting to see if higher temperature data might follow the same scaling with a larger $T^*$. In recent pump probe experiment on CeCoIn$_5$ \cite{Qi2020}, two collective modes have been detected to emerge below these two temperatures, respectively. It could be possible that there exist multiple hybridization processes and the Nernst effect is more sensitive to the lower temperature one. In any case, the Nernst coefficient seems very different from the Hall coefficient. The latter is mostly dominated by the local moment contribution, while our observed scaling here indicates that the local moment contribution is largely suppressed in the Nernst effect, presumably by the energy derivative in Eq.~(\ref{eq2}). In this respect, the enhanced Nernst signal is primarily associated with the asymmetry of the quasiparticle density of states (as a result of hybridization) rather than that of the scattering rate. 

\begin{figure}[t]
\begin{center}
\includegraphics[width=0.45\textwidth]{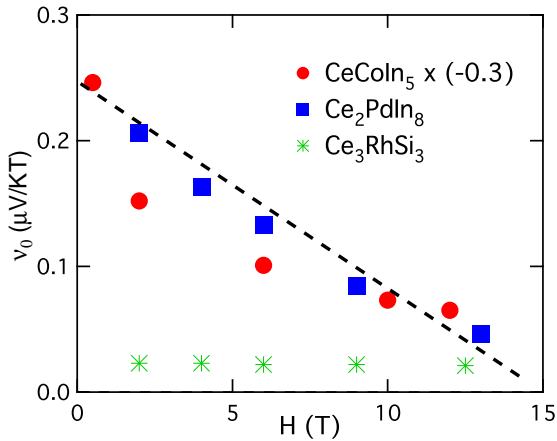}
\caption{The derived value of $\nu_0$ as a function of the magnetic field for CeCoIn$_5$, Ce$_2$PdIn$_8$, and Ce$_3$RhSi$_3$. The data for CeCoIn$_5$ is multiplied by -0.3. The dashed line is a guide to the eye.}
\label{fig3}
\end{center}
\end{figure}

Second, the magnitude of the Nernst signal is tentatively correlated with the value of $T^*$. The latter sets roughly the effective bandwidth of the heavy quasiparticle spectra \cite{Mo2012} and is inversely proportional to the density of states and its slope at the Fermi energy \cite{Yang2012}. In Fig.~\ref{fig1}, a large $\nu_0$ as in CeCoIn$_5$ or Ce$_2$PdIn$_8$ seems quite typically to have a small $T^*$, while all other compounds have relatively larger $T^*$ and smaller $\nu_0$. Indeed, it has been shown previously that the Nernst coefficient in CeCoIn$_5$ is correlated with the quasiparticle effective mass or the specific heat \cite{Behnia2009}, both of which are inversely proportional to $T^*$ \cite{Yang2012}. It is known that $T^*$ is typically the order of 10-100 K, only a few hundredths of the bandwidth of conduction electrons in a normal metal. Hence the large Nernst signal in heavy fermion materials is primarily associated with their small quasiparticle bandwidth. Clearly, such a correlation is not exact and the Nernst signal may vary a lot owing to the complicated variation of ground state properties. The scaling is supposed to be interrupted by these low temperature orders \cite{Yang2012,Shirer2012}. Figure~\ref{fig3} compares the derived $\nu_0$ as a function of the magnetic field for CeCoIn$_5$, Ce$_2$PdIn$_8$ and Ce$_3$RhSi$_3$. We see no common trend of $\nu_0$ versus $H$ in three compounds. For Ce$_3$RhSi$_3$, $\nu_0$ is almost field independent, possibly associated with its ferromagnetic property. For CeCoIn$_5$ and Ce$_2$PdIn$_8$, $\nu_0$ decreases monotonically with increasing field, reflecting the suppression of heavy quasiparticles at high field. On the other hand, if we look closer into the data of CeCoIn$_5$, there appears to be a change of slope around the quantum critical point at 4.1 T \cite{Zaum2011,Yang2014}, but more field data will be needed to clarify this possibility.

\begin{figure}[t]
\begin{center}
\includegraphics[width=0.45\textwidth]{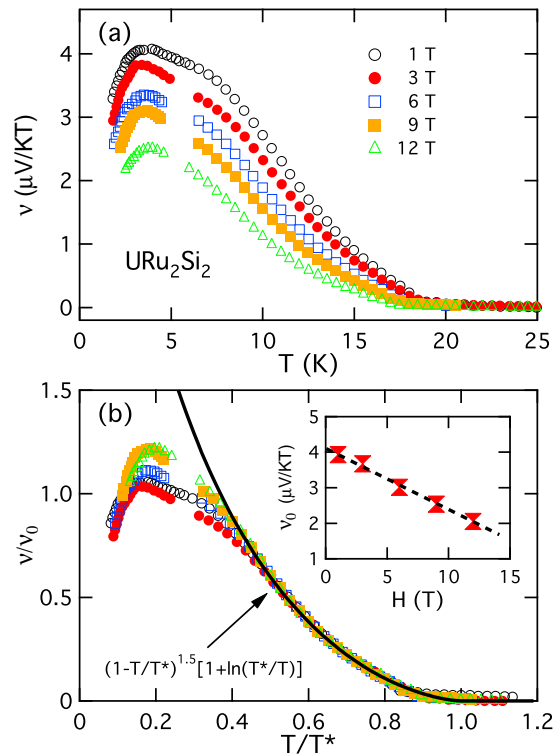}
\caption{(a) The Nernst coefficient of URu$_2$Si$_2$ under different magnetic fields reproduced from the literature \cite{Bel2004a}. (b) The universal scaling of the normalized Nernst data as a function of the dimensionless temperature. The value of $T^*$ varies slightly from about 22 to 17 K with increasing field. The solid line is the predicted scaling formula of emergent heavy quasiparticles in the two-fluid model \cite{Yang2008}. The inset shows the derived $\nu_0$ decreasing linearly with field. The dashed line is a guide to the eye.}
\label{fig4}
\end{center}
\end{figure}

\section{V. Application to the hidden order state of U$\text{Ru}_2\text{Si}_2$}
With these in mind, we extend our analysis to the mysterious heavy fermion compound URu$_2$Si$_2$. Intriguingly, a giant Nernst signal has been observed in the hidden order phase below $17.5 $ K \cite{Bel2004a} rather than the coherence temperature of about 55 K \cite{Yang2008b}. Thermopower and Hall measurements point up the possibility of a small Fermi energy, a low carrier density, and a long scattering time \cite{Bel2004a}. Other more exotic scenarios involve the presence of chiral orders or chiral fluctuations \cite{Kotetes2010,Sumiyoshi2014,Yamashita2015}. With increasing magnetic field, the Nernst signal shows successive anomalies, suggesting a series of Fermi surface reconstructions \cite{Pourret2013}. But still, as shown in Fig.~\ref{fig4}, all data can be scaled to collapse on the same universal curve with a derived $T^*$ of 17-22 K, just around the hidden order temperature and depending weakly on the magnetic field. Again, our data is limited below 25 K. It will be interesting to see if the normal state data follow the above scaling with a higher $T^*$. On the other hand, the validity of the same scaling in the hidden order phase seems to suggest that the hidden order also possesses certain aspect of hybridization physics \cite{Balatsky2011}, among many other dazzling theories \cite{Mydosh2011}. This is in agreement with the point contact measurement \cite{Park2012}, in which a significant Fano resonance has been observed in the hidden order phase, suggesting the presence of hybridization \cite{Yang2009}. Previously, it has also been shown in experiment that a band of heavy quasiparticles drops below the Fermi level across the transition \cite{Syro2009}, which probably causes the change in the Nernst effect. With increasing field, the derived $\nu_0$ decreases linearly as shown in the inset of Fig.~\ref{fig4}(b), indicating a diminishing hybridization in accordance with the suppression of the hidden order  \cite{Knafo2018}. 

\section{VI. Summary and conclusion}

Putting together, the Nernst effect in heavy fermion materials may be typically categorized into three regimes, a high temperature regime with an almost constant background, a wide intermediate temperature regime with enhanced signal and universal temperature dependence due to emergent heavy quasiparticles with a narrow bandwidth, and a low temperature regime where the Nernst coefficient tends to be suppressed. Similar suppression is also present in the Hall data \cite{Nakajima2007}. While the high temperature constant comes from conduction electrons or $f$ valence bands (as in YbAl$_3$), the low temperature limit varies a lot depending on the Fermi liquid or other ordered states. In between, one has the universal scaling and may ideally expect a continuing increase near a quantum critical point as observed in Ce$_3$RhSi$_3$ at low field. It might be helpful to note that the scaling of the Nernst coefficient is highly nontrivial. It results from the unique prediction of the two-fluid model combined with the skew scattering formalism and the special energy derivative form of Eq.~(\ref{eq2}). The latter eliminates the contribution of the local moment component that is typically material dependent and dominates most other physical quantities and, as a result, exposes the universal temperature scaling of the emergent heavy quasiparticles. This explains why universality has not been widely observed in other transport and thermodynamic properties in heavy fermion materials. We note that our derivation is by no means exact and may be limited by the validity of the skew scattering mechanism. The skew scattering formula has been confirmed experimentally to dominate the Hall coefficient in most heavy fermion compounds  \cite{Schoenes1988,Sakamoto2003,Chen2004,Sugawara2005,Paschen2005,Kohler2007}, but may suffer from slight derivation below the coherence temperature and become most violated inside some ordered states or the Fermi liquid, where both the Hall and Nernst effects are governed by well-defined Landau quasiparticles. The scaling formula is hence expected to break down at very low temperatures, as also observed in Fig.~\ref{fig2}, and a different formula should be applied. To date, there is still no satisfactory microscopic theories for the Nernst effect in heavy fermion materials. Our observed scaling provides a unified basis and a potential clue for its future investigations.

\section*{ACKNOWLEDGEMENTS}

This work was supported by the National Key R\&D Program of China (Grant No. 2017YFA0303103) and the National Natural Science Foundation of China (NSFC Grant No. 11774401, No. 11974397).

\end{document}